\documentclass[10pt,a4paper]{article}
\usepackage{amsmath,amssymb,graphicx,geometry,indentfirst,textgreek,natbib,authblk,aas_macros}
\begin{document}
\title{\textbf{Cosmic-Ray Convection-Diffusion Anisotropy}}
\author[1]{Yiran Zhang\footnote{zhangyr@pmo.ac.cn}}
\author[2]{Siming Liu\footnote{liusm@swjtu.edu.cn}}
\author[1]{Dejin Wu\footnote{djwu@pmo.ac.cn}}
\affil[1]{Key Laboratory of Planetary Sciences, Purple Mountain Observatory, Chinese Academy of Sciences, Nanjing 210023, China}
\affil[2]{School of Physical Science and Technology, Southwest Jiaotong University, Chengdu 610031, China}
\maketitle
\begin{abstract}
Under nonuniform convection, the distribution of diffusive particles can exhibit dipole and quadrupole anisotropy induced by the fluid inertial and shear force, respectively. These convection-related anisotropies, unlike the Compton-Getting effect, typically increase with the cosmic-ray (CR) energy, and are thus candidate contributors for the CR anisotropy. In consideration of the inertial effect, CR observational data can be used to set an upper limit on the average acceleration of the local interstellar medium in the equatorial plane to be on the order of 100 $ \mu \text{m}/\text{s}^2 $. Using Oort constants, the quadrupole anisotropy above 200 TeV may be modeled with the shear effect arising from the Galactic differential rotation.

\end{abstract}
\section{Introduction}
The arrival direction of cosmic rays (CRs) has been studied with a long history. Observations show that CRs have a power-law spectrum with a highly isotropic angular distribution, whose relative amplitude of anisotropy is on the order of 0.1\% at TeV energies. The anisotropy mainly increases with the CR energy, while observations in recent years reveal a local decrease in about 10--100 TeV, and correspondingly a flip of the dipole anisotropy phase roughly from the Galactic anti-center to center direction \citep{2016ApJ...826..220A,2017ApJ...836..153A,2018ApJ...861...93B}. For ultrahigh-energy CRs (UHECRs, i.e., those above EeV), the anisotropy seems to increase to the order of 1\%, with the dipole phase returning to the anti-center direction \citep{2017Sci...357.1266P,2020ApJ...898L..28A}.

It is recognized that the CR anisotropy mainly arises from diffusion rather than convection. Reference is generally made to the point that a standard diffusion model of particles just predicts a dipole anisotropy proportional to the diffusion coefficient, which typically increases with the particle energy. On the contrary, the dipole anisotropy induced by uniform convection, i.e., the Compton-Getting \citep[CG;][]{1935PhRv...47..817C} effect, is energy-independent for a power-law spectrum of particles, and is therefore inconsistent with the CR observation.

Obviously, the above argument is not based on a complete description of the convection-diffusion picture, which can only be studied in the concept of the general fluid rest frame with the fluctuation-relaxation theory. \cite{1988ApJ...331L..91E} proposed the Bhatnagar-Gross-Krook (BGK) approximation of such a problem to study the CR viscosity. The fluctuational anisotropy obtained in the work includes additional effects from nonuniform convection, which are typically proportional to the diffusion coefficient, implying that they have the potential to explain the CR observation. However, these effects seem to be overlooked in the CR anisotropy problem. This paper provides some further discussion about such effects in combination with the observation.
\section{Multipole Expansion}
Generally, the multipole expansion of the relative anisotropy can be written as
\begin{align}
\frac{\Delta f}{f}=-\boldsymbol{D} \cdot \frac{\boldsymbol{p}}{p}+Q_{ij}\frac{p^ip^j}{2p^2},\label{me}
\end{align}
where $ \boldsymbol{D} $ and $ Q_{ij} $ are the dipole and quadrupole moment with respect to the observer's line-of-sight (LOS) vector $ -\boldsymbol{p}/p $, respectively, and $ \boldsymbol{p} $ is the particle momentum.

In a scattering system without external forces, the anisotropic fluctuation $ \Delta f $ \citep[Eq.~(5) in][]{1988ApJ...331L..91E} of the particle phase-space distribution function in a local reference frame in which the scattering centers are at rest satisfies the multipole expansion, with
\begin{align}
\boldsymbol{D}&=\tau \left( v\boldsymbol{\nabla }\ln f-\frac{\boldsymbol{a}}{v}\frac{\partial \ln f}{\partial \ln p} \right) ,\label{dm}\\
Q_{ij}&=2\tau S_{ij}\frac{\partial \ln f}{\partial \ln p},\label{qm}
\end{align}
where $ f $ is the isotropic distribution function, $ v $ is the particle speed, $ \tau $ is the scattering relaxation time, and
\begin{align}
\boldsymbol{a}&=\dot{\boldsymbol{u}},\\
S_{ij}&=\frac{1}{2}\left( \frac{\partial u_j}{\partial x^i}+\frac{\partial u_i}{\partial x^j} \right) -\frac{\delta _{ij}}{3}\boldsymbol{\nabla }\cdot \boldsymbol{u}
\end{align}
are the fluid acceleration and shear-rate tensor, respectively, with $ \boldsymbol{u} $ the non-relativistic flow velocity, $ x^i $ the $ i $th spatial coordinate, and $ \delta _{ij} $ the Kronecker delta. As seen, the above anisotropies do not depend directly on $ \boldsymbol{u} $ itself, but on its derivative. Via an inertial transformation, one obtains the CG anisotropy with the dipole moment $ \left( \boldsymbol{u}/v \right) \partial \ln f/\partial \ln p $ for $ u\ll v $ \citep{2017PrPNP..94..184A}, which is determined directly by $ \boldsymbol{u} $. For CRs, $ \boldsymbol{u} $ may be considered primarily as the Alfv\'en velocity.

A brief interpretation to the convection-diffusion approximation can be as follows. In the rest frame of a scattering center, the scattered particle ``forgets'' its initial state of motion in the scattering time. In a collection of the scattering centers, this may be translated to some extent as ``a given anisotropy $ \Delta f $ can completely be relaxed in a time of $ \tau $ only in the fluid rest frame''. This reference frame is generally non-inertial, with the inertial force $ -\boldsymbol{a}p/v $ and shear-restoring force $ -S_{ij}p^j $ acting on a particle, in response to the change of $ \boldsymbol{u} $. A pure scattering system can be considered to be free of external forces. To isotropize the distribution in the time $ \tau $, $ \Delta f $ should be the distributional increment caused by the time reversal motion, i.e., by changes of phase-space coordinates in a time of $ -\tau $, during which the Liouville's theorem is valid. Thus if $ \tau $ is small, one can write $ \Delta f\approx -\tau \left( \boldsymbol{v}\cdot \boldsymbol{\nabla }f+\dot{\boldsymbol{p}}\cdot \partial f/\partial \boldsymbol{p} \right) $. This directly yields Eq.~(\ref{me}) when $ \dot{\boldsymbol{p}} $ represents the inertial plus shear force. The spatial gradient term corresponds to the standard diffuse anisotropy, while for convenience we shall refer to that associated with the inertial and shear effect as the inertial and shear anisotropy, respectively. Note that $ \Delta f $ can consistently be observed under fluctuation-relaxation equilibrium.
\section{Inertial Anisotropy}
For a microscopic steady flow, the magnitude of the inertial anisotropy in Eq.~(\ref{dm}) is generally $ O\left( v/u \right) $ times smaller than that of the shear anisotropy Eq.~(\ref{qm}). For turbulence, the shear as well as CG effect will vanish after ensemble averaging over some scales on which the flow is fully stochastic, i.e., $ \left< \boldsymbol{u}\right> =\boldsymbol{0} $, while the inertial effect can exist as $ \left< \boldsymbol{a} \right> =\left< \boldsymbol{u}\cdot \boldsymbol{\nabla u} \right> \ne \boldsymbol{0} $. If the fluid is further incompressible, i.e., $ \boldsymbol{\nabla}\cdot \boldsymbol{u}=0 $, one has $ \left< \boldsymbol{a} \right> =2u\boldsymbol{\nabla}u/3 $. That is to say, over sufficiently large (space-time) scales, the average anisotropy associated with the background flow should be dominated by the inertial effect, with the dipole direction roughly along $ \boldsymbol{\nabla}u $ for $ \partial f/\partial p<0 $. CR observations do include some average effects due to time integration, thus anisotropies on $ O\left( u/v \right) $ may be suppressed to some extent in the data.

\begin{figure}
	\centering
	\includegraphics[width=0.5\textwidth]{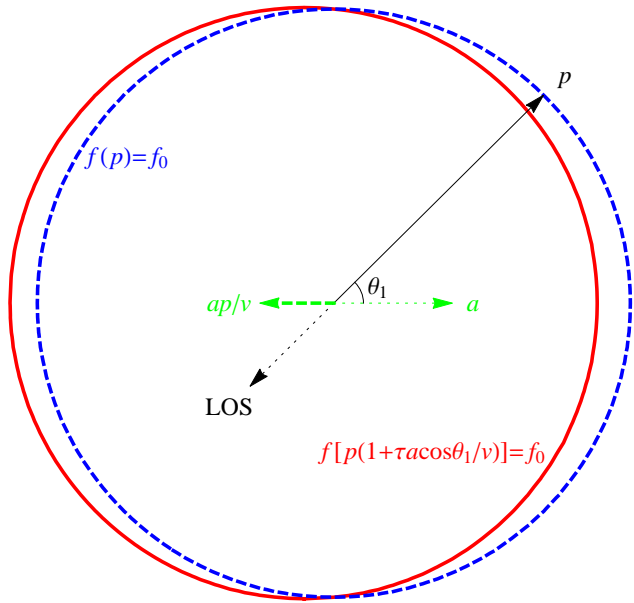}
	\caption{Schematic view of the inertial dipole anisotropy, where $ \nu =1 $, $ \mathcal{A}_1=0.3 $. The red solid curve is an isoline of the anisotropic distribution of particles in the momentum space in the $ a $-$ p $ plane. The anisotropy can be relaxed in a time of $ \tau $ ($ \ll v/a $) due to the inertial force, which is shown with the green thick dashed arrow, to produce an isotropic distribution whose contour is the blue dashed circle.}\label{f1}
\end{figure}

In general, $ \tau $ is an increasing function of $ p $, and is related to the diffusion coefficient $ \kappa $ in the Parker's transport equation by $ \kappa =\tau v^2/3 $ \citep{1988ApJ...331L..91E}. Thus not only diffusion but also nonuniform convection is eligible for modeling the CR anisotropy, which mainly increases with the particle energy. The inertial term in Eq.~(\ref{me}) can be written as
\begin{align}
-\frac{\nu +2}{\left| \nu +2 \right|}\mathcal{A}_1\cos \theta _1,
\end{align}
where $ \nu =-2-\partial \ln f/\partial \ln p $ is the (ultra-relativistic) energy spectral index, $ \theta _1 $ is the angle between $ \boldsymbol{p} $ and $ \boldsymbol{a} $, and
\begin{align}
\mathcal{A}_1=\frac{\tau a}{v}\left| \nu +2 \right| \approx 5.6\times 10^{-4}\left( \frac{c}{v} \right) ^{3}\frac{\kappa}{10^{29}\text{ cm}^2/\text{s}}\frac{a}{100\ \mu \text{m}/\text{s}^2}\frac{\left| \nu +2 \right|}{5}\label{ia}
\end{align}
is the first-harmonic amplitude with $ c $ the speed of light. On this order of magnitude, if there is any observable contribution to the CR anisotropy by the inertial effect, it is likely to be related to inner structures of the local interstellar medium (ISM). For comparison, we emphasize that the acceleration of the local standard of rest on Galactic scales is only about 0.2 $ \text{nm}/\text{s}^2 $ \citep{2021A&A...649A...9G}. Fig.~\ref{f1} is the schematic view of the inertial anisotropy.

\begin{figure}
	\centering
	\includegraphics[width=1\textwidth]{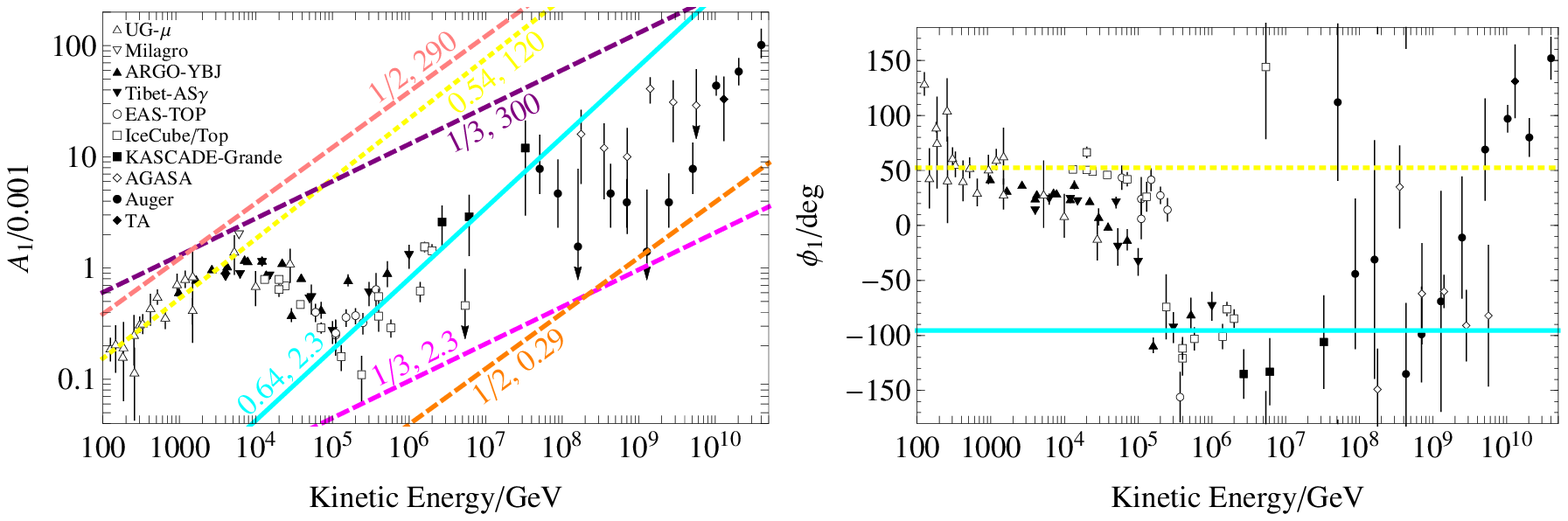}
	\caption{Comparison of the dipole anisotropy in RA reported by CR observations and that calculated with the inertial effect Eq.~(\ref{ia}) assuming a full-sky RA projection, where $ \nu =\left( 2.7+3.1 \right) /2 $. The colored pair of numbers represents $ \left[ \partial \ln \kappa /\partial \ln p,\kappa \left( \text{TeV} \right) a\cos \delta _1/\left( 10^{29}\text{ cm}^2\mu \text{m}/\text{s}^3 \right) \right] $. Long-dashed lines indicate limiting cases in which the inertial anisotropy and $ A_1 $ data can overlap. The yellow short-dashed and cyan solid line show the fit to data below 5 TeV and in 200 TeV--50 PeV, respectively. The data are from underground muon detectors \citep[UG-$ \mu $ data in][]{2005ApJ...626L..29A}, Milagro \citep{2009ApJ...698.2121A}, ARGO-YBJ \citep{2015ApJ...809...90B,2018ApJ...861...93B}, Tibet-AS$ \gamma $ \citep{2005ApJ...626L..29A,2017ApJ...836..153A}, EAS-TOP \citep{1995ICRC....2..800A,1996ApJ...470..501A,2009ApJ...692L.130A}, IceCube/Top \citep{2010ApJ...718L.194A,2012ApJ...746...33A,2013ApJ...765...55A,2016ApJ...826..220A}, KASCADE-Grande \citep{2019ApJ...870...91A}, AGASA \citep{1999APh....10..303H}, Auger \citep{2020ApJ...891..142A} and TA \citep{2020ApJ...898L..28A}.}\label{f2}
\end{figure}

Within a classical fluid concept in which the flow property is independent of the particle energy, the inertial effect cannot solely explain the CR anisotropy at all energies, nor can the typical scenario of pure diffusion on Galactic scales (e.g., the leaky-box model) provide a complete explanation. It has been shown that the observed amplitude dip and phase flip of the TeV dipole anisotropy may be ascribed to a superposition effect of some source components \citep{2016PhRvL.117o1103A,2019JCAP...12..007Q,2022MNRAS.511.6218Z}. Since the diffuse component is presumably essential for the dipole moment, it seems difficult to strictly quantify the CR inertial effect. Nevertheless, one can still estimate an upper limit of the local ISM acceleration by considering that the anisotropy produced by Eq.~(\ref{ia}) should not much exceed the observed value. On the other hand, a significant inertial effect on the CR anisotropy observationally requires a minimum acceleration. In conclusion, for the inertial anisotropy ($ \nu \sim 3 $) to overlap the amplitude data in Fig.~\ref{f2}, we have
\begin{align}
1\ \mu \text{m}/\text{s}^2\lesssim \frac{\kappa \left( \text{TeV} \right)}{10^{29}\text{ cm}^2/\text{s}}a\cos \delta _1\lesssim 100\ \mu \text{m}/\text{s}^2,\label{ar}
\end{align}
where $ \delta _1 $ is the declination (Dec) of the dipole direction. This constrains the average acceleration only in the equatorial plane, because here we consider only RA projected anisotropy data, which are most reported by existing observations. Such data can be seen as fitting results of the Dec-averaged RA-projected relative fluctuation
\begin{align}
\psi =\frac{1}{\sin \delta _{\text{up}}-\sin \delta _{\text{low}}}\int_{\delta _{\text{low}}}^{\delta _{\text{up}}}{\frac{\Delta f}{f}\cos \delta \text{d}\delta},\label{ep}
\end{align}
where $ \delta _{\text{low}} $ and $ \delta _{\text{up}} $ are the lower- and upper-limit Dec of the observatory’s time-integrated field of view \citep{2017PrPNP..94..184A}. For the dipole component, one has $ \psi _1=A_1\cos \left( \alpha -\phi _1 \right) $, where $ \alpha $ is the RA of the LOS direction, and the phase $ \phi _1 $ is equal to the RA of the dipole direction $ \alpha _1 $. For our crude estimation, it may be enough to assume a full-sky scan with $ A_1=\pi \mathcal{A}_1\cos \delta _1/4 $.

Although it is possible to neutralize an overproduced inertial anisotropy via introducing diffuse components with specified dipole directions, the upper limit in Eq.~(\ref{ar}) makes sense for physical simplicity. Say, if the CR anisotropy below 10 TeV is ascribed to the inertial effect, the data above 10 TeV are difficult to be explained. Moreover, as there are indications that the dipole anisotropy around EeV has a similar phase-flip behavior to that around 100 TeV, which needs to be further verified by future precise observations of UHECRs, all data below EeV are unlikely to have a simple explanation by an inertial effect with energy-independent $ \boldsymbol{a} $. However, for turbulence, the flow property depends on the scale of view. Since this scale must be related to the Larmor radius, $ \boldsymbol{a} $ can have dependence on the particle energy. It is then possible to model the anisotropy decrease via a competition effect of low- and high-energy flows with opposite directions of $ \boldsymbol{a} $. The observed anisotropy then can be used to probe the properties of the turbulent flow.
\section{Shear Anisotropy}
As we know, any deformation of a continuous medium can be decomposed into an isotropic part, i.e., an expansion (or compression) characterized by the strain-rate tensor $ \delta _{ij}\boldsymbol{\nabla }\cdot \boldsymbol{u}/3 $, and an anisotropic constant-volume part known as a pure shear with the strain rate $ S_{ij} $ \citep{1959thel.book.....L}. Interestingly, the shear anisotropy of the microscopic distribution corresponds exactly to that of the macroscopic medium via Eq.~(\ref{qm}), while the isotropic deformation affects only the isotropic part of the particle distribution through the adiabatic process.

\begin{figure}
	\centering
	\includegraphics[width=1\textwidth]{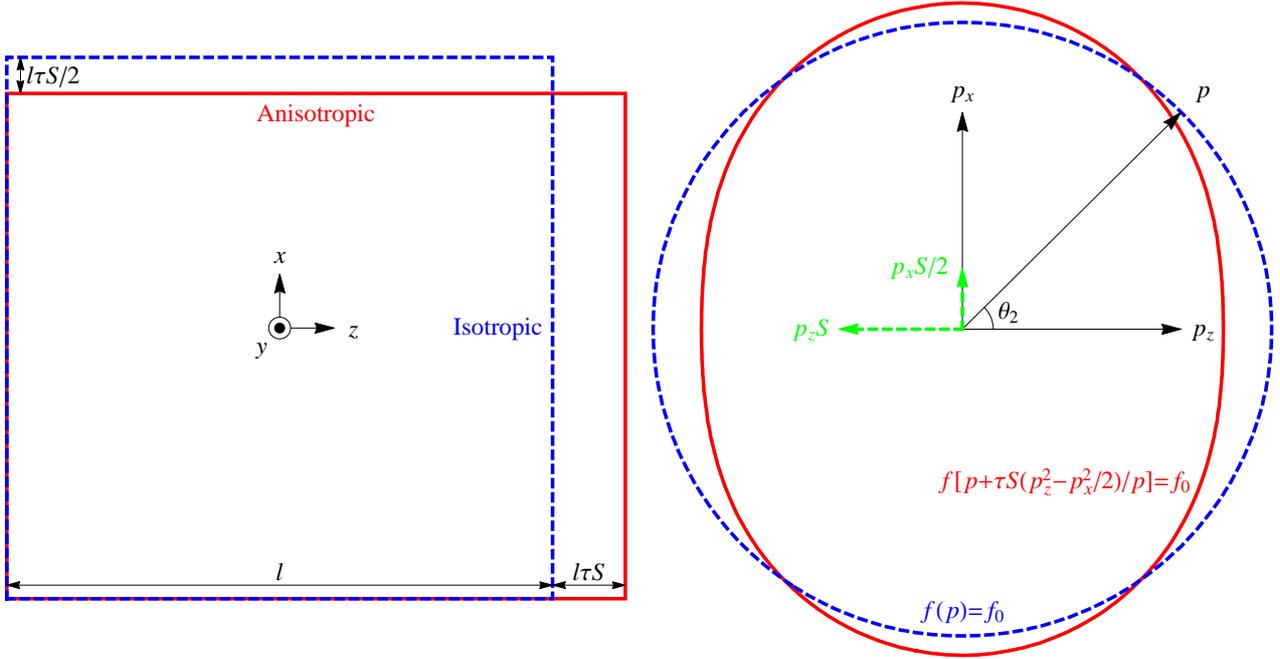}
	\caption{Schematic view of the simple extension. Left: Fluid deformation in a time of $ \tau $ ($ \ll 1/S $). The blue dashed-line square is the projection of an isotropic (3D) fluid element on the $ x $-$ z $ plane. The red solid-line rectangle shows the element after constant-volume lengthening along $ \hat{\boldsymbol{z}} $. Right: Contours of particle momentum-space distributions (with $ p_y=0 $) corresponding to deformation states in the left-hand panel, where $ \nu =1 $, $ \mathcal{A}_2=0.3 $. Green dashed arrows represent shear-restoring forces on a particle.}\label{f3}
\end{figure}

As a traceless symmetric tensor, $ S_{ij} $ generally contains five independent parameters, and can (locally) be diagonalized via a rotation transformation to the eigenbasis. The diagonalized $ S_{ij} $ is expressed as a sum of shear-rate tensors that represent simple extensions (or shortenings) along the eigenvectors, with the parameters being three Euler angles (determining directions of the eigenvectors) and two of the eigenvalues. A simple extension can be obtained from a one-dimensional (1D) flow, i.e., $ \boldsymbol{u}=u_z\left( z \right) \hat{\boldsymbol{z}} $, where $ u_z $ spatially depends only on $ z $, and $ \hat{\boldsymbol{z}} $ denotes the unit vector in the $ z $ direction. In this system, $ S_{ij} $ contains three independent parameters, i.e., two of the Euler angles (determining $ \hat{\boldsymbol{z}} $) and one of the eigenvalues. Then the quadrupole term in Eq.~(\ref{me}) can be described via the Legendre polynomial
\begin{align}
-\frac{4}{3}\frac{S_{zz}}{\left| S_{zz} \right|}\frac{\nu +2}{\left| \nu +2 \right|}\mathcal{A}_2\text{P}_2\left( \cos \theta _2 \right) ,
\end{align}
where $ \theta _2 $ is the angle between $ \boldsymbol{p} $ and $ \hat{\boldsymbol{z}} $, and
\begin{align}
\mathcal{A}_2=\frac{3}{4}\tau \left| S_{zz} \right| \left| \nu +2 \right| \approx 4\times 10^{-4}\left( \frac{c}{v} \right) ^{2}\frac{\kappa}{10^{29}\text{ cm}^2/\text{s}}\frac{\left| S_{zz} \right|}{10\text{ Myr}^{-1}}\frac{\left| \nu +2 \right|}{5}\label{sa}
\end{align}
is the second-harmonic amplitude. Fig.~\ref{f3} is the schematic view of the simple extension, showing that the particle distribution tends to increase in shortening directions of the fluid element (if $ \nu >-2 $). For a steady state, the bulk acceleration is
\begin{align}
u_z\frac{\partial u_z}{\partial z}=\frac{3}{2}u_zS_{zz}\approx 4.8\frac{u_z}{10\text{ km}/\text{s}}\frac{S_{zz}}{\text{10 Myr}^{-1}}\text{ nm}/\text{s}^2.
\end{align}
This means that the quadrupole anisotropy, compared with the dipole, is much more easily induced by a nonuniform flow. The presence of such a quadrupole effect in CR data depends on the interpretation of observations. As mentioned previously, the shear anisotropy may be important only if regularity of the flow survives after ensemble averaging. Nevertheless, there is likely to be a significant effect of the ISM shear on the CR anisotropy, as the characteristic acceleration producing the observed quadrupole amplitude is only around $ \text{nm}/\text{s}^2 $.

To compare with observations over a broad energy range, it is better to study the RA projected anisotropy. Considering a full-sky scan for simplicity, following Eq.~(\ref{ep}) the projection of any quadrupole fluctuation onto the equatorial plane is
\begin{align}
\psi _2&=\frac{1}{4}\int_{-\frac{\pi}{2}}^{\frac{\pi}{2}}{\left( \begin{matrix}
	\cos \delta \cos \alpha\\
	\cos \delta \sin \alpha\\
	\sin \delta
	\end{matrix} \right) ^{\text{T}}\left( \begin{matrix}
	Q_{\xi \xi}&		Q_{\xi \eta}&		Q_{\xi \zeta}\\
	Q_{\xi \eta}&		Q_{\eta \eta}&		Q_{\eta \zeta}\\
	Q_{\xi \zeta}&		Q_{\eta \zeta}&		Q_{\zeta \zeta}
	\end{matrix} \right) \left( \begin{matrix}
	\cos \delta \cos \alpha\\
	\cos \delta \sin \alpha\\
	\sin \delta
	\end{matrix} \right) \cos \delta \text{d}\delta}\notag \\
&=\frac{1}{3}\left[ \frac{Q_{\xi \xi}-Q_{\eta \eta}}{2}\cos \left( 2\alpha \right) +Q_{\xi \eta}\sin \left( 2\alpha \right) \right] =A_2\cos \left[2\left( \alpha -\phi _2\right)\right],
\end{align}
where $ \xi $, $ \eta $ and $ \zeta $ are equatorial rectangular coordinates, with $ \xi $ and $ \zeta $ denoting the vernal-equinox and north-polar direction, respectively. That is to say, ideally the quadrupole effect on the equatorial plane can always be described via second Fourier harmonics. To express $ A_2 $ and $ \phi _2 $ in terms of eigenvalues of $ Q_{ij} $ (with the eigenvectors in $ x $-$ y $-$ z $ directions), we rewrite the equatorial representation of $ Q_{ij} $ as
\begin{align}
R_{\alpha _2}^{\zeta}R_{\frac{\pi}{2}-\delta _2}^{\eta}R_{\varphi}^{\zeta}\left( \begin{matrix}
Q_{xx}&		0&		0\\
0&		Q_{yy}&		0\\
0&		0&		Q_{zz}
\end{matrix} \right) \left( R_{\alpha _2}^{\zeta}R_{\frac{\pi}{2}-\delta _2}^{\eta}R_{\varphi}^{\zeta} \right) ^{\text{T}},
\end{align}
where $ R_{\theta}^i $ denotes the rotation matrix rotating a vector by an angle of $ \theta $ about the $ i $ axis, $ \alpha _2 $ and $ \delta _2 $ are the RA and Dec of $ \hat{\boldsymbol{z}} $, respectively, and $ \varphi $ is the azimuthal parameter in the $ x $-$ y $ plane. After some calculations, we find
\begin{align}
A_2\left[ \begin{matrix}
\cos \left( 2\phi _2\right)\\
\sin \left( 2\phi _2\right)\\
\end{matrix} \right] =\frac{R_{2\alpha _2}}{2}\left[ \begin{matrix}
\frac{Q_{xx}-Q_{yy}}{3}\left( 1-\frac{\cos ^2\delta _2}{2} \right) \cos \left( 2\varphi \right) +\frac{Q_{zz}}{2}\cos ^2\delta _2\\
\frac{Q_{xx}-Q_{yy}}{3}\sin \delta _2\sin \left( 2\varphi \right)\\
\end{matrix} \right] ,\label{psa}
\end{align}
where $ R_{\theta} $ is the 2D rotation matrix. 

\begin{figure}
	\centering
	\includegraphics[width=1\textwidth]{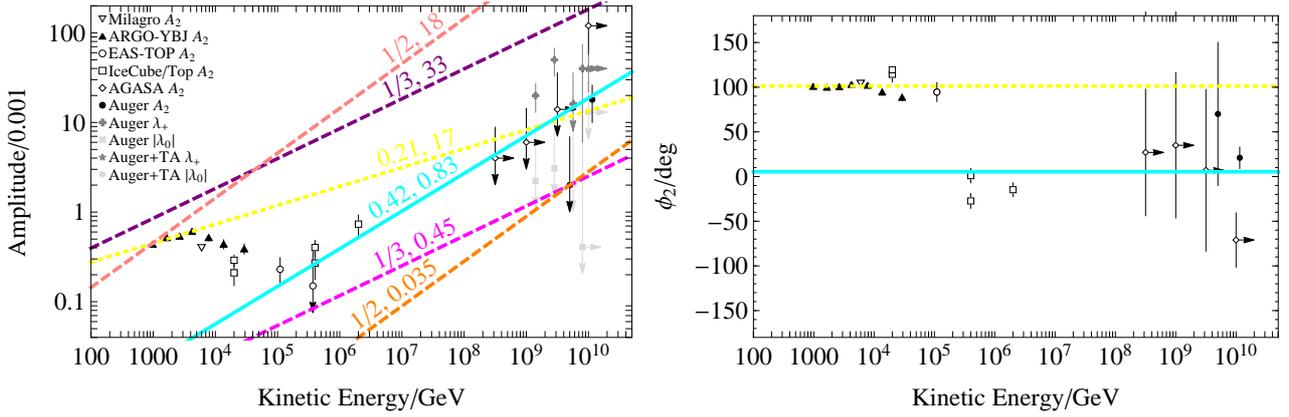}
	\caption{Comparison of the quadrupole anisotropy reported by CR observations and that calculated with the 1D shear model Eq.~(\ref{sa}) assuming a full-sky RA projection, where $ \nu =\left( 2.7+3.1 \right) /2 $. The colored pair of numbers represents $ \left[ \partial \ln \kappa /\partial \ln p,\kappa \left( \text{TeV} \right) \left| S_{zz}\right| \cos ^2\delta _2/\left( 10^{29}\text{ Myr}^{-1}\text{cm}^2/\text{s} \right)  \right] $. Long-dashed lines indicate limiting cases in which the shear anisotropy and $ A_2 $ data can overlap. The yellow short-dashed and cyan solid line show the fit to the $ A_2 $ data below 5 TeV and above 200 TeV, respectively. The data are from Milagro \citep{2009ApJ...698.2121A}, ARGO-YBJ \citep{2015ApJ...809...90B}, EAS-TOP \citep{2009ApJ...692L.130A}, IceCube/Top \citep{2010ApJ...718L.194A,2012ApJ...746...33A,2013ApJ...765...55A}, AGASA \citep{1999APh....10..303H}, Auger \citep{2012ApJS..203...34P,2018ApJ...868....4A} and Auger+TA \citep{2014ApJ...794..172A}.}\label{f4}
\end{figure}

In particular, for a 1D flow with $ Q_{xx}=Q_{yy} $, one has $ A_2=2\mathcal{A}_2\cos ^2\delta _2/3 $, $ \phi _2=\alpha _2+n\pi $ if $ Q_{zz}>0 $, and $ \phi _2=\alpha _2+\left( n+1/2 \right) \pi $ if $ Q_{zz}<0 $, where $ n $ is an arbitrary integer. We may naively make use of this simple formalism, just as we did to obtain Eq.~(\ref{ar}), to give an order-of-magnitude estimate of the ISM shear rate required to produce the CR anisotropy shown in Fig.~\ref{f4}. The data suggest that
\begin{align}
0.1\text{ Myr}^{-1}\lesssim \frac{\kappa \left( \text{TeV} \right)}{10^{29}\text{ cm}^2/\text{s}}\left| S_{zz}\right| \cos ^2\delta _2\lesssim 10\text{ Myr}^{-1},
\end{align}
and there is also an amplitude dip and phase flip for the second-harmonic signal around 100 TeV. Note that we have shifted some of the $ \phi _2 $ data reported by original literature by $ \pm \pi $ to obtain the two-phase-like shape, based on the fact that all values of $ \phi _2+n\pi $ are equivalent. Despite the lack of observations, the available data show no strong indication for a local decrease of $ A_2 $ at ultrahigh energies, in agreement with constant $ \phi _2 $ in the same energy range. Only then it is possible to explain the RA projected quadrupole anisotropy above 200 TeV with an energy-independent fluid shear, e.g., the 1D flow with $ \alpha _2\sim 0^{\circ} $ and $ Q_{zz}>0 $. The phase flip around 100 TeV may indicate a transition between two quadrupole anisotropy regimes. A possible scenario is that low-energy CRs are largely affected by the local interstellar magnetic field (LIMF), whose turbulence gives rise to multipole anisotropies \citep{2014PhRvL.112b1101A,2017PrPNP..94..184A}. For a $ \mu $G LIMF, if its coherence length is 0.1 pc, 100 TeV should be the critical energy above which protons are no longer trapped by the LIMF. Intuitively, it is also possible that the phase flip arises from a competition effect of energy-dependent flows, or of magnetic fields averaged over small and large scales.

A few observations also report detailed parameters of $ Q_{ij} $ from fits directly to the CR anisotropy sky map. As far as we know, above 200 TeV such data have only been reported by the Pierre Auger Observatory and Telescope Array at EeV energies \citep{2012ApJS..203...34P,2014ApJ...794..172A,2018ApJ...868....4A}. The reported eigenvalues, i.e., $ \lambda _+ $, $ \lambda _- $ and $ \lambda _0 $ ($ \lambda \equiv Q/2 $) shown in Fig.~\ref{f4}, reveal an interesting feature: $ \lambda _+\sim -\lambda _- \gg \left| \lambda _0 \right| $. This motivates us to set $ Q_{zz}=2\lambda _0=0 $, which indicates $ A_2\leqslant 2\lambda _+/3 $ according to Eq.~(\ref{psa}). Since the $ A_2 $ and $ \lambda _+ $ data appear to be comparable, $ \delta _2 $ should be far from $ 0^{\circ} $ if $ \varphi =\left( n+1/2\right) \pi /2 $. The angular parameters are unreported in \cite{2012ApJS..203...34P}, while above 10 EeV they can be found in \cite{2014ApJ...794..172A}. \cite{2018ApJ...868....4A} reported all independent components of $ Q_{ij} $ above 4 EeV in equatorial rectangular coordinates. However, these reconstruction results are still not statistically significant.

In the shear anisotropy model, the Auger quadrupole data imply $ S_{zz}=0 $, which however does not represent a (quasi) 1D flow. Instead, the simplest configuration may be an incompressible simple-shear flow, i.e., $ \boldsymbol{u}=u_Y\left( X \right) \hat{\boldsymbol{Y}} $ with the following diagonalization of $ S_{ij} $,
\begin{align}
\left( \begin{matrix}
0&		S_{XY}&		0\\
S_{XY}&		0&		0\\
0&		0&		0
\end{matrix} \right) R_{\frac{\pi}{4}}^{z}=R_{\frac{\pi}{4}}^{z}\left( \begin{matrix}
S_{XY}&		0&		0\\
0&		-S_{XY}&		0\\
0&		0&		0
\end{matrix} \right) ,
\end{align}
where
\begin{align}
\frac{1}{2}\left| \frac{\partial u_Y}{\partial X} \right|=\left| S_{XY} \right|=\frac{\lambda _+}{\tau \left| \nu +2 \right|}\approx 19\left( \frac{v}{c}\right) ^2\frac{10^{33}\text{ cm}^2/\text{s}}{\kappa}\frac{5}{\left| \nu +2 \right|}\frac{\lambda _+}{0.01}\text{ Gyr}^{-1}.
\end{align}
For suitable values of $ \kappa $, $ \nu $ and $ \lambda_ + $, this rate of shear can be comparable with the Oort constant $ A $, which measures the local azimuthal shear rate in the Galactic plane, and is significantly greater than the radial shear rate $ C $ and divergence rate $ K $. We thus naturally speculate that the azimuthal shear due to the Galactic differential rotation has significant effects on the CR quadrupole anisotropy. The weak CG effect does indicate a Galactic co-rotation of CRs \citep{2006Sci...314..439A}, which implies high electric conductivity such that the magnetic field is frozen into the rotating plasma fluid.

\begin{figure}
	\centering
	\includegraphics[width=1\textwidth]{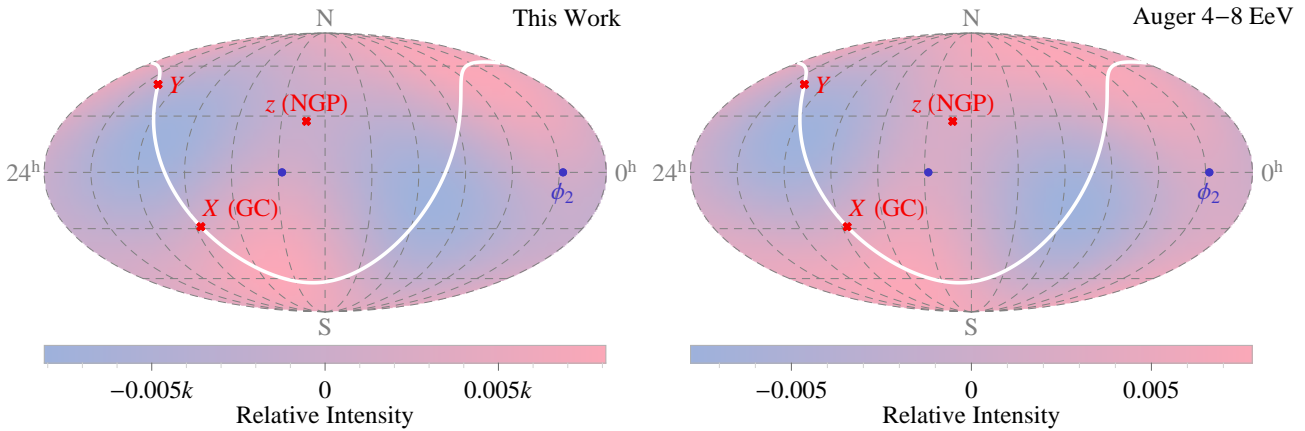}
	\caption{Left: Equatorial sky map of the CR anisotropic intensity (relative to the isotropic part) induced by the Galactic differential rotation with the Oort shear rate Eq.~(\ref{oc}), where $ \nu =\left( 3.2+2.6 \right) /2 $, $ \kappa /k=10^{33}\text{ cm}^2/\text{s} $. The white solid curve shows the Galactic disk. The Galactic rectangular frame is indicated with red crosses (NGP: north Galactic pole). Blue dots mark directions of peaks of the RA projected intensity, with $ \phi _2\approx 28^{\circ} $. Right: Most similar intensity profile to the left-hand panel, derived with reconstructed $ Q_{ij} $ in \cite{2018ApJ...868....4A}, where the parameters used are given in Tab.~\ref{t1}.}\label{f5}
\end{figure}

Assuming vertical homogeneity of the local ISM with respect to the Galactic plane (i.e., $ \partial \boldsymbol{u}/\partial z=\boldsymbol{\nabla}u_z=\boldsymbol{0} $), the ``realistic'' average $ S_{ij} $ in the Galactic coordinate system (with $ \hat{\boldsymbol{X}} $ and $ \hat{\boldsymbol{Y}} $ directed at the Galactic $ 0^\circ $ and $ 90^\circ $ longitude, respectively) is expressed as follows in terms of the Oort constants \citep{2019ApJ...872..205L},
\begin{align}
\left( \begin{matrix}
\frac{K}{3}+C&		A&		0\\
A&		\frac{K}{3}-C&		0\\
0&		0&		-\frac{2}{3}K
\end{matrix} \right) \approx \left( \begin{matrix}
-3.3&		15&		0\\
15&		2.2&		0\\
0&		0&		1.1
\end{matrix} \right) \text{ Gyr}^{-1}.\label{oc}
\end{align}
This can directly be used to calculate the CR shear anisotropy. Eliminating all LOS-independent and dipole terms in Eq.~(\ref{me}), the relative quadrupole intensity with excess-deficit symmetry is obtained. The left-hand panel of Fig.~\ref{f5} shows the predicted shear anisotropy sky map, whose full-sky RA projection gives
\begin{align}
A_2\approx 0.003\frac{\kappa}{10^{33}\text{ cm}^2/\text{s}}\label{a2}
\end{align}
and $ \phi _2\approx 28^{\circ} $. The right-hand panel of Fig.~\ref{f5} shows the most similar intensity profile within uncertainties, to the left-hand panel, derived with reconstructed $ Q_{ij} $ in 4--8 EeV (at a median energy of 5 EeV) reported by Auger \citep{2018ApJ...868....4A}, with the parameters adopted given in Tab.~\ref{t1}. For $ \kappa \left( 5 \text{ EeV} \right) \sim 10^{33}\text{ cm}^2/\text{s} $, which is roughly consistent with some theoretical expectations of UHECR propagation \citep{2012JCAP...07..031G}, the two maps are highly coincident. It should be noted that within uncertainties the Auger profile can also differ remarkably from the model prediction. On the other hand, the inertial anisotropy calculated from Eq.~(\ref{ia}) using the Oort constants gives $ \mathcal{A}_1\left( 5 \text{ EeV} \right) \sim 10^{-5} $, which can be neglected compared with the observed CR dipole anisotropy. However, the true average inertial effect may not be so weak as it depends on the variance of turbulence.
\begin{table}
	\centering
	\begin{tabular}{c|cc}
		& \cite{2018ApJ...868....4A} & Fig.~\ref{f5} Right-hand Panel\\
		\hline
		$ Q_{\zeta \zeta} $ & $ -0.01\pm 0.04 $ & $ 0.011 $\\
		$ Q_{\xi \xi}-Q_{\eta \eta} $ & $ -0.007\pm 0.029  $ & $ 0.0097 $\\
		$ Q_{\xi \eta} $ & $ 0.004\pm 0.015  $ & $ 0.007 $\\
		$ Q_{\xi \zeta} $ & $ -0.02\pm 0.019 $ & $ -0.001 $\\
		$ Q_{\eta \zeta} $ & $ -0.005\pm 0.019 $ & $ 0.0093 $
	\end{tabular}
	\caption{Reconstructed $ Q_{ij} $ in 4--8 EeV reported by \cite{2018ApJ...868....4A}, and that used in the right-hand panel of Fig.~\ref{f5}.}\label{t1}
\end{table}

Given the large mean free path $ 3\kappa \left( \text{5\ EeV} \right) /v\sim 30 $ kpc, the fluid description of UHECRs in fact makes sense on intergalactic scales. The coincidence of the Oort and Auger anisotropy may imply a co-rotation of the intergalactic medium with the Galactic disk. Recent observations of X-ray absorption have inferred that within about 100 kpc from the Galactic center (GC), the hot gaseous halo spins at comparable velocities as the disk \citep{2016ApJ...822...21H}. As $ \tau \left( \text{5\ EeV} \right) \sim 100\text{ kyr}\ll 70\text{ Myr}\approx 1/A $, the UHECR relaxation is still rapid in comparison with the quasi-static shear flow.
	
After all, the Oort shear rate Eq.~(\ref{oc}) is derived typically from stellar and ISM populations within a heliocentric distance of a few kpc \citep{2017MNRAS.468L..63B}. Such ``disk'' hydrodynamics should be established under $ \kappa \lesssim 10^{32}\text{ cm}^2/\text{s} $, which constrains $ A_2\lesssim 3\times 10^{-4} $ according to Eq.~(\ref{a2}). Also given the change of $ \phi _2 $ around 200 TeV, the Oort anisotropy model is better to be compared with PeV CR data, yet detailed results of $ Q_{ij} $ at such energies have not been reported in literature. Nevertheless, the RA projected data in Fig.~\ref{f4} imply $ A_2\left( \text{PeV} \right) \sim 3\times 10^{-4} $. We may then define that the Oort anisotropy is important if $ \kappa \left( \text{PeV} \right) \gtrsim 10^{31}\text{ cm}^2/\text{s} $, which can be reached via $ \partial \ln \kappa \left( \text{TeV--PeV} \right) /\partial \ln p\gtrsim 1/2 $, provided that $ \kappa \left( \text{TeV} \right) \sim 3\times 10^{29}\text{ cm}^2/\text{s} $ as inferred from recent studies of AMS-02 data \citep{2020PhRvD.101j3035D}. The Oort anisotropy dominates the PeV range if the validity of Eq.~(\ref{oc}) can be extended to intergalactic hydrodynamics with $ \kappa \left( \text{PeV} \right) \sim 10^{32}\text{ cm}^2/\text{s} $. In that case, the post-knee quadrupole anisotropy may be an indirect evidence of dark halo spin.
\section{Summary}
In this paper, based on the BGK description of the convection-diffusion system, we discuss the effect of nonuniform convection on the CR anisotropy. Assuming that relaxation of particle trajectories can only be observed in the fluid rest frame, the inertial and shear-restoring force give rise to dipole and quadrupole anisotropy in the particle distribution, respectively. Unlike the CG effect, these two convection-related anisotropies typically increase with the CR energy, and are thus eligible for modeling the CR observation. After looking into the data, we conclude:

(a) There is no explanation of the CR anisotropy at all energies simply by energy-independent nonuniform convection;

(b) The decrease of the dipole anisotropy in 10--100 TeV implies an upper limit about 100 $ \mu \text{m}/\text{s}^2 $ for the local ISM average acceleration in the equatorial plane;

(c) The quadrupole anisotropy above 200 TeV may be relevant to the shear effect due to the Galactic differential rotation (including halo spin), which can partially be characterized with Oort constants.

We hope that future experiments such as LHAASO \citep{2022ChPhC..46c0001M} can provide more precise measurements and comprehensive analyses of CR multipole anisotropies above PeV, which are essential for further validation of the model.
\section*{Acknowledgment}
This work is partially supported by grants from the National Natural Science Foundation of China (grant No.~U1931204), National Key R\&D Program of China (2018YFA0404203), and Jiangsu Funding Program for Excellent Postdoctoral Talent (2022ZB476).
\bibliographystyle{aasjournal}
\bibliography{ref}
\end{document}